\documentclass{emulateapj}
\usepackage{apjfonts}
\newcommand{\lsun}{\rm L$_\odot$}

\shorttitle{HD 15115 at 1.1\micron} 
\shortauthors{Debes et al.}

\begin{document}
\title{Color Gradients Detected in the HD~15115 Circumstellar Disk}
\author{John H. Debes\altaffilmark{1},Alycia J. Weinberger\altaffilmark{1}, Inseok Song\altaffilmark{2}}

\altaffiltext{1}{Department of Terrestrial Magnetism, Carnegie Institution of 
Washington, Washington, DC 20015 (debes@dtm.ciw.edu)}
\altaffiltext{2}{Department of Physics and Astronomy, University of Georgia, Athens, Georgia 30605}

\begin{abstract}
We report HST/NICMOS coronagraphic images of the HD 15115 circumstellar disk at 1.1\micron.  We find a similar morphology to that seen in the visible and at H band--an edge-on disk that is asymmetric in surface brightness.  Several aspects of 
the 1.1\micron\ data are different, highlighting the need for multi-wavelength images of each circumstellar disk.  We find a flattening to the western surface brightness profile at 1.1\micron\ interior to 2\arcsec\ (90~AU) and a warp in the western half of the disk.  We measure the surface brightness profiles of the two disk lobes and create a measure
of the dust scattering efficiency between 0.55-1.65\micron\ at 1\arcsec, 2\arcsec, and 3\arcsec.  At 2\arcsec\ the western lobe has a neutral spectrum up to 1.1\micron\ and a strong absorption or blue spectrum $>$1.1\micron, while a blue trend is 
seen in the eastern lobe.  At 1\arcsec\ the disk has a red F110W-H color in both lobes.
\end{abstract}

\keywords{stars:individual(HD 15115)--circumstellar matter}

\section{Introduction}
Raw material from the interstellar medium is processed through circumstellar disks into many different types of planetary bodies.  Debris disks are a useful environment to study the composition of planetesimals that form around stars different from our own Sun.  Multi-wavelength scattered light observations of disks provide one avenue for determining the compositon of dust caused by collisions of planetesimals.

HD~15115 possesses one of a growing number of spatially resolved debris disks amenable to multiwavelength observations.  HD 15115 is an F2 star at a distance of 45$\pm$1~~pc \citep{vanl} and was observed to have an IR excess \citep{silverstone00}.  Unresolved emission from its disk was detected at 60, 100, and 850 \micron, implying a 
ring at $\sim$35~AU with a temperature of 62~K \citep{zuckerman04,decin03,williams06}.  Recently, \citet[][hereafter KFG07]{kalas07} discovered circumstellar emission from a combination of Hubble Space Telescope/Advanced Camera for Surveys (HST/ACS) and Keck near-IR AO coronagraphic imaging.  Surface brightness (SB) profiles were reported for V and H bands.  Furthermore, they observed an extreme size and brightness difference between the two lobes of the disk and shape asymmetries around the disk midplane similar to that seen around $\beta$ Pictoris and indicative of a second disk at 
a different inclination \citep{kalas95,golimowski06}.  We have imaged HD 15115 at 1.1\micron\ with HST; we present our observations in \S\ref{s:obs}, and analyze the results in \S\ref{s:analysis}.  We discuss the implications of our work in \S\ref{s:conc}

\section{Observations}
\label{s:obs}
We observed
HD~15115 with the Hubble Space Telescope's Near-Infrared
and Multi-Object Spectrometer (NICMOS) \citep{thompson98} 
using the coronagraph in Camera 2.  We
used the F110W filter with a central wavelength of 1.1\micron (0.8\micron-1.4\micron). 
 Observations were taken on 2006 October 20
for HD~15115 and HD~16647, the PSF reference.  

\begin{figure}
\plotone{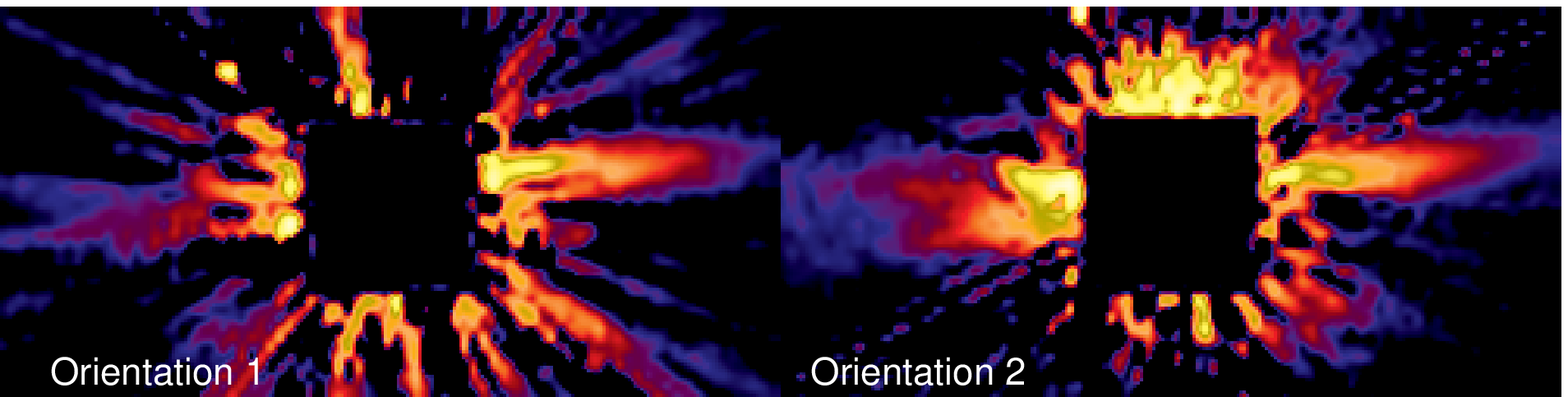}
\plotone{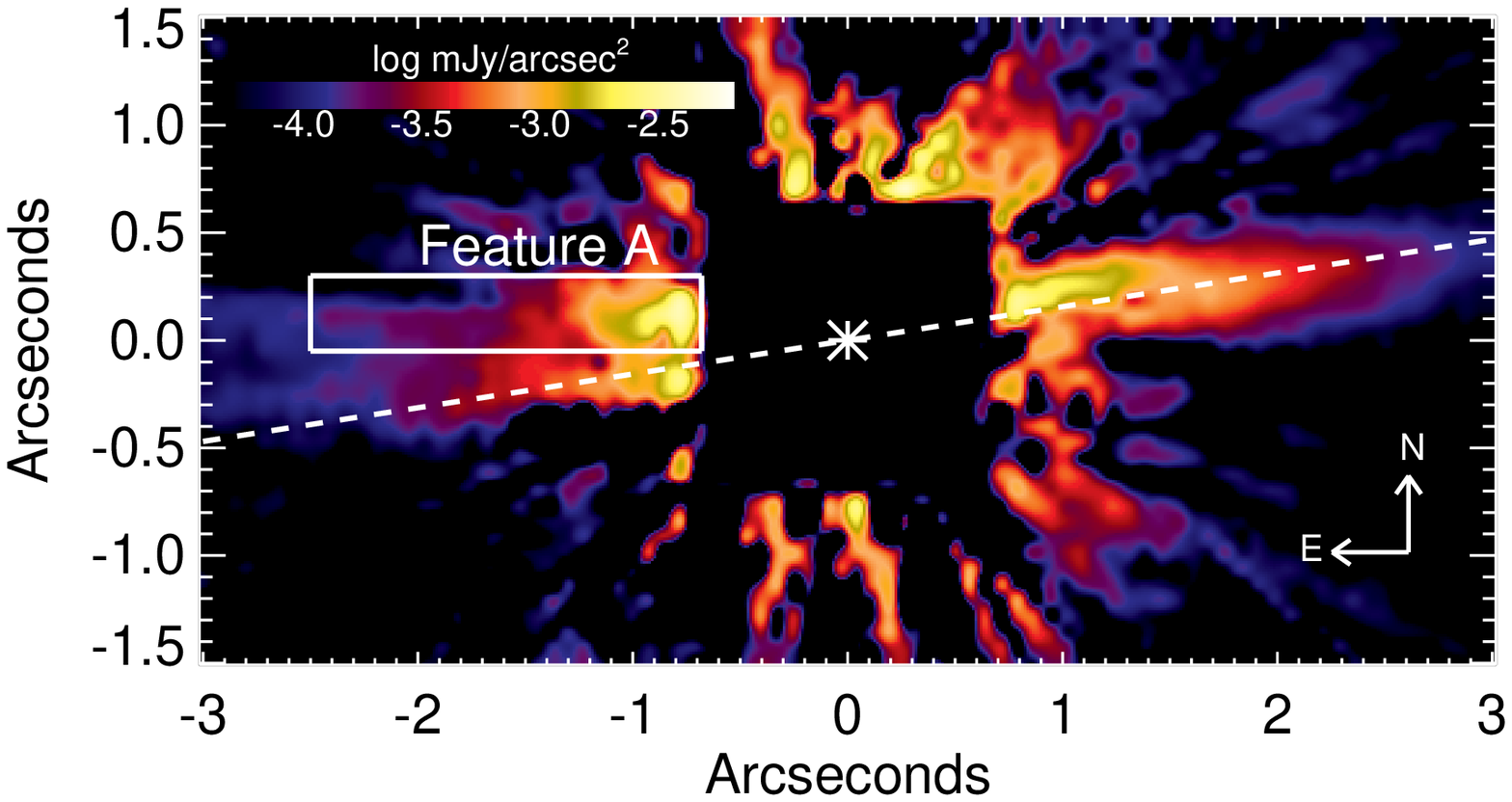}
\caption{\label{fig:final}(top) False color images of the observations of HD~15115 in the F110W filter at two spacecraft orientations.  The images are logarithmically scaled.
(bottom)False color, final combined image of the HD~15115 disk, showing the contaminating feature A.  The dashed line represents the nominal position angle of the disk at 278.9$^\circ$, while the asterisk represents the position of the
star.  We have masked out separations in the disk within 0\farcs68, where PSF subtraction residuals dominate.  Noticeable are a warp on the western lobe, and 
the asymmetries between the western and eastern lobes.}
\end{figure}

Accurate disk photometry requires a correct scaling ratio between the target star and its PSF reference.  Both stars would saturate the detector in the shortest exposure.  Therefore, we estimated the F110W flux density for both stars with synthetic photometry using
 the CALCPHOT task of the SYNPHOT\footnote[1]{http://www.stsci.edu/resources/software\_hardware/stsdas} package in conjunction with Kurucz models tabulated in the
SYNPHOT library.  We found 
a best fit match to HD~15115's luminosity and effective temperature
based on its parallax from the Hipparcos satellite, its Tycho-2 $B$ and $V$
photometry, and 2MASS $J$, $H$, and $K_s$ photometry.  The model assumed a solar metallicity and the best fit was for a $\log{g}$=5.0, T$_{eff}$=6750~K star, 
with a luminosity of 3.11\lsun.  We fit solar metallicity models to HD 16647, finding a best fit for a $\log{g}$=5.0 T$_{eff}$=6750~K star and a luminosity of 4.01\lsun.  The resulting SYNPHOT scaling in F110W is 0.600.  Based on previous experience, \citep[][for example]{debes08}, SYNPHOT is accurate to $\sim$3\%, which we will adopt as our {\em a priori} estimate of the scaling uncertainty.
Coronagraphic images of the disk were taken at two spacecraft orientations.  
Th images were subtracted, registered, and combined following the procedures of \citet{debes08}.  The individual subtracted and combined images are shown in Figure \ref{fig:final}. 

\section{Analysis}
\label{s:analysis}
Figure \ref{fig:final} shows that the disk is detected at the same position angle (PA) as KFG07, despite the presence of one strong residual streak close to the position of the eastern lobe of the disk (Feature A in Figure \ref{fig:final}).  Perpendicular to the disk, strong residuals also exist.  We measured the PA, full-width at half-maximum (FWHM) of the disk midplane, and SB profiles as a function of radius from
the central star.

\subsection{Disk Geometry}
\begin{figure}
\plotone{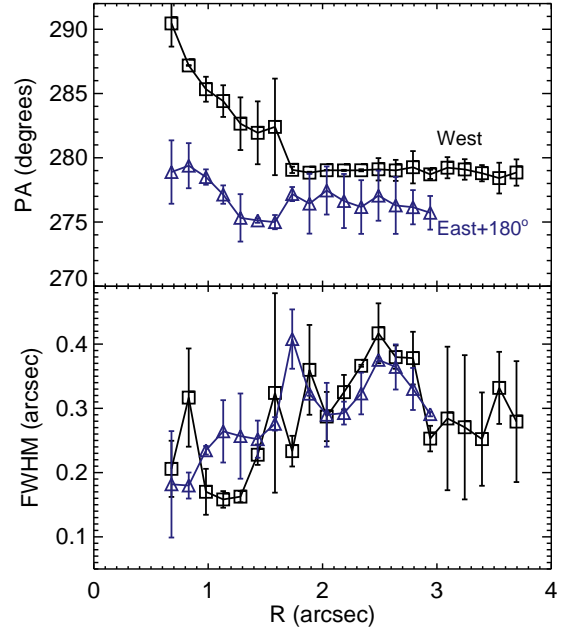}
\caption{\label{fig:westmeas}Measurements of the HD~15115 disk midplane PA (top)
and FWHM (bottom) as a function of distance for the two lobes of the disk.  Squares represent the western lobe, while triangles represent the eastern lobe.  The sharp rise in PA for the western lobe interior 1\farcs7 is due to the warp, and the Eastern lobe's PA beyond 1\farcs8 is marginally different than the Western lobe.}
\end{figure}

We measured the PA and FWHM of the western lobe of the disk from 0\farcs68 (30~AU) to 3\farcs69 (165~AU).  For the eastern lobe we took measurements from 0\farcs68 to 3\farcs0 (135~AU).  In our analysis we rotated the final disk images so that the disk midplane was oriented along an image row.   We took
3$\times$13 (0\farcs23$\times$0\farcs98) pixel cuts of the disk centered on the brightest pixel of the disk and summed in a direction perpendicular to the midplane direction.  We then fit a Gaussian curve to the disk emission to measure the location of the midplane (and thus the local PA) as well as the FWHM of the disk.  We took an average of the values for both orientations and estimated the uncertainty as being half the difference between the two measurements.  Figure \ref{fig:westmeas} shows the PA and FWHM of the western lobe of the disk as a function of radius.  Beyond 1\farcs7, we find a PA for the disk of 278.9$^\circ\pm0.2^\circ$, which is consistent with KFG07's value of 278.5$^\circ\pm0.5^\circ$.  However, interior to 1\farcs7, we note that the disk PA increases with decreasing distance
from the star to a maximum PA of 290$^\circ \pm 1^\circ$ at 0\farcs68.  The changing PA
of the midplane is seen at both orientations.  Furthermore, we find at 2\arcsec, a FWHM of 0\farcs29$\pm$0.04  which is consistent with the FWHM of 0\farcs19$\pm$0.1 reported by KFG07.  The median FWHM of the disk from 0\farcs68-3\farcs69 is 0\farcs26$\pm$0\farcs07.  The FWHM at 1.1\micron\ is broadened by the NICMOS F110W PSF, which has a FWHM of 0\farcs1, implying an intrinsic FWHM for the disk midplane of $\sim$0\farcs24 or 11~AU.

Measurement of the eastern side of the disk is complicated by the contamination of Feature A from PSF subtraction.  To measure the FWHM and PA, we masked out pixels heavily contaminated by this feature.  We again took the average of the two sets of measurements and estimated the uncertainties as being half the difference between the two measurements.  We find no evidence of a warp in the eastern lobe, which has a median PA of 96.6$^\circ\pm1.3^\circ$.  To compare to the PA of the Western side we add 180$^\circ$ to get 276.6$^\circ$, consistent with no significant bowing in the large scale structure of the disk as is seen
for HD~61005 \citep{hines07}.  The median value of the eastern FWHM is 0\farcs29$\pm$0\farcs06.

\subsection{Surface Brightness Profiles}

\begin{figure}
\plotone{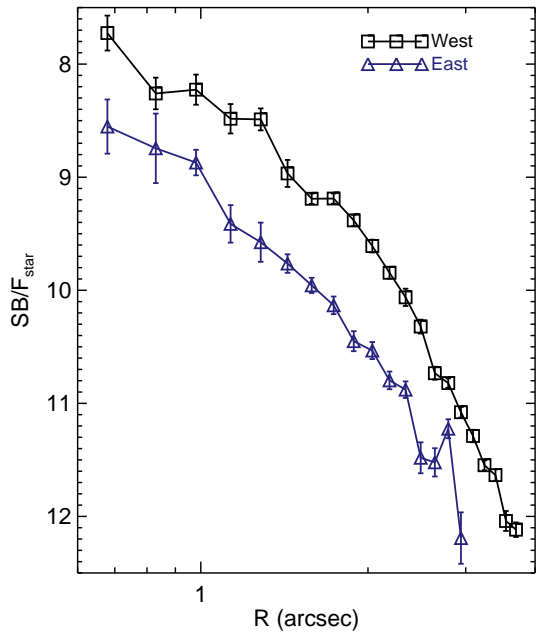}
\caption{\label{fig:sbprofile}The surface brightness profiles of the eastern and western lobes of the disk in F110W magnitudes per square arcsecond for a 0\farcs23$\times$0\farcs23 aperture.  No corrections were applied.  The symbols are the same as in Figure \ref{fig:westmeas}}
\end{figure}

Figure \ref{fig:sbprofile} shows a comparison of the 1.1\micron\ SB profiles in $\Delta m_{F110W}$ arcsec$^{-2}$ relative to the central star ($m_{F110W}$=6.12) for the eastern and western lobes of the disk by taking 0\farcs23$\times$0\farcs23 apertures centered on the midplane of the disk.  
Between
0.7-1.8\arcsec, the disk SB of the western lobe falls as $r^{-1.4\pm0.1}$, while at radii $>$1.8\arcsec, the SB drops as $r^{-3.56\pm0.06}$.  The eastern lobe drops steadily as $r^{-2.14\pm0.06}$ from 0\farcs7-3\arcsec.

\subsection{Scattering Efficiency vs. Wavelength}

Combining our data with the reported V and H SB data, a measure of the scattering efficiency of the dust around HD 15115 can be constructed.  The measured SB at any radius in the disk can be self-consistently
modeled for composition and grain size distribution provided a knowledge of the density distribution of the dust is known \citep[see details in][]{debes08}.   The SB of a dust disk at a particular radius is $\propto F_\star \phi(\theta)Q_{sca}\frac{L_\star}{L_{IR}}$, where $\phi(\theta)$ is the phase function of the dust at a particular scattering angle, $Q_{sca}$ is the scattering efficiency of the dust, and $\frac{L_\star}{L_{IR}}$ is the ratio of stellar luminosity to the IR luminosity of the dust distribution.  This is difficult for HD 15115 because the density distribution of the dust is not known through, for example, spatially resolved thermal emission images.  However, multi-wavelength imaging can provide useful compositional information even in the absence of spatially resolved thermal emission constraints.  In the case of HD 15115, three images at different wavelengths will be insufficient to provide a definitive picture, but can constrain certain compositions.

In order to compare results across different wavelengths, instruments, and telescopes, it is imperative that a careful and consistent treatment of SB measurements is implemented.  Observations of disks at different wavelengths can be affected by the respective PSFs of telescopes with different diffraction limits and Strehl ratios, but no standard technique is applied to the study of circumstellar disks.  Intuitively,
one would expect minor corrections as a function of wavelength provided that the disk structure did not significantly change between wavelengths and that the disk itself was mostly resolved.  KFG07 estimated aperture corrections using point source
PSFs.  Other methods have included deconvolution of the data \citep{golimowski06}, or using apertures that capture the majority of the observed emission from the disk
at each wavelength \citep{schneider99,weinberger02,debes08}.   We modeled the effect of the different PSFs of NICMOS, ACS, and Keck H-band adaptive optics on the measured surface brightness for the 0\farcs23$\times$0\farcs23 aperture. 

Our approach to calculating the corrections derives from the reliable theoretical PSFs generated by TinyTim\footnote{http://www.stsci.edu/software/tinytim/} for HST as well as using a Keck H PSF corresponding to the October observations of HD~15115 (M. Fitzgerald, personal communication).  We took the PSFs for each instrument and  convolved them with model disks
constructed from the measured power-law slopes and disk FWHMs for each observation.  We then measured the SB as a function of distance for the original and convolved models with the 0\farcs23$\times$0\farcs23 aperture in F110W and a 0\farcs25$\times$0\farcs25 aperture for V and H and took the ratio as a single valued multiplicative correction to the observed SB profiles.  We verified that the slope of the surface brightness profiles did not change due to convolution.  To estimate the uncertainty in this correction, we used the standard deviation of the measurements between 2\arcsec-3\arcsec,
the region where all the observations overlap.  For the ACS coronagraphic PSF, we found
a correction of 1.23$\pm$0.02, for the H band we found a correction of 1.56$\pm$0.06, and for F110W we found a correction of 1.36$\pm$0.04.

To compare the NICMOS and other measurements, we must make a correction for the slightly different photometric aperture sizes of 0\farcs23$\times$0\farcs23 and 0\farcs25$\times$0\farcs25 respectively.  To do this we oversampled our model
 NICMOS images by a factor of 3 and calculated the predicted difference in SB measurements between our 0\farcs23$\times$0\farcs23 aperture and a 0\farcs25$\times$0\farcs25 one.  This multiplicative correction changes with distance from the star, and ranges from 1.2 to 0.8.

\begin{figure}
\plotone{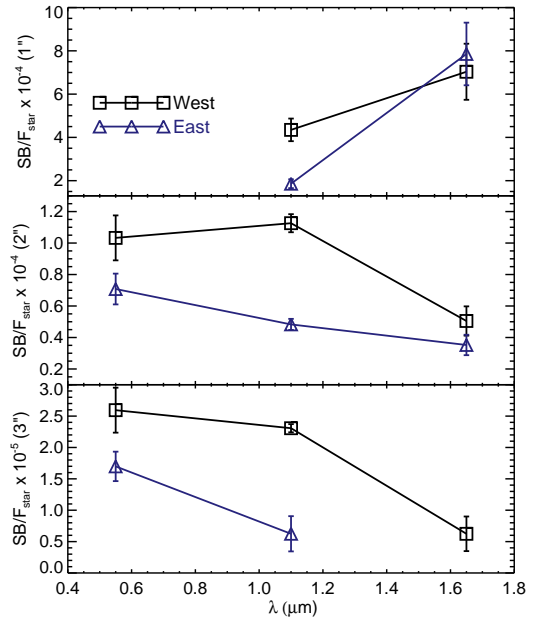}
\caption{\label{fig:spec} Comparison of the scattering efficiency (or color) vs.wavelength of HD~15115's disk at 1\arcsec (45~AU), 2\arcsec (90~AU), and 3\arcsec (135~AU).  The symbols are the same as in Figure \ref{fig:westmeas}.  The disk dramatically changes color over wavelength and distance from the central star.}
\end{figure}

In Figure \ref{fig:spec} we plot SB/F$_{\nu,\star}$ as a function of wavelength for HD 15115's two disk lobes at 1\arcsec, 2\arcsec, and 3\arcsec.  This observed quantity is $\propto Q_{sca}$.  Our 1.1\micron\ data provides a third point in the measure of HD 15115's scattering efficency when combined with the 0.55\micron\ (V) and 1.65\micron\ (H) measurements in KFG07.  We derive the uncertainties in the scattering efficiency for 0.55\micron\ and 1.65\micron\ based on
Figure 3 in KFG07.  Where no error bars were given for the 1.65\micron\ data in KFG07, we assumed an uncertainty of 20\%.  In the western lobe at 2\arcsec, we find that the disk's scattering efficiency is neutral out to 1.1\micron\ but then drops such that the
spectrum is blue at 1.65\micron.  The spectrum is the same 
from 0.55-1.65\micron\ at 3\arcsec.  At 2\arcsec, the eastern lobe is strongly
blue and at 3\arcsec\ the blue trend continues out to 1.1\micron.  KFG07 did not report SB for the H data at 3\arcsec, but if the spectrum is the same as the inner part of the disk, one would expect the blue trend to continue to longer wavelengths.

Finally, interior to about 1\farcs8, we can get a rough sense of the near-IR colors of the disk in the east and west.  We took KFG07's measurements of the two lobes in H at 1\arcsec and compared them to our data.  In both cases the disk is red, the result of the difference in SB profiles between F110W and H interior to 2\arcsec.

\section{Discussion}
\label{s:conc}

The new features we have observed in the disk at 1.1\micron\ add questions to the nature of the disk and the dust.  The increase in PA towards the inner part of the disk could be caused by a genuine warp, or perhaps the presence of a second disk inclined at $>$12$^\circ$ from the main disk
as is seen in $\beta$-Pictoris.  This is supported by the detection of an asymmetry in the northern part of the disk seen in the ACS and H data of KFG07. 

KFG07 noted an overall blue V-H color for the Western lobe of the disk.  The presence of a neutral scattering efficiency out to 1.1\micron\ as well as a red F110W-H color interior to 2\arcsec\ suggests that the spectrum of the HD~15115 disk is more complicated than the V and H data suggest.  The abrupt change in scattering efficiency beyond 2\arcsec\ from F110W to H may be due to the presence of strong absorption at $\sim$1.65\micron.  One candidate would be strong absorption due to water ice, something not yet detected in
the scattered light of a disk.  If that is the case, even deeper absorption should be present at 2 and 3.5\micron.
 Two possibilities exist to explain the red F110W-H color of the inner disk--either there is evidence of strong absorption due to olivine (which has an absorption feature at 1\micron), or the presence of a very red material in the disk, possibly similar to what is seen in HD~100546 \citep{ardila07} or HR~4796A \citep{debes08}.

The age of HD~15115 is not well constrained, which would be helpful in understanding the origins of the debris disk.  Kinematically, it is marginally consistent with membership in the 12~Myr-old $\beta$-Pictoris Moving Group (BPMG) \citep{moor06}.  However, this is not borne out by backtracking the position of HD~15115 based on its proper motion and radial velocity using the methods of \citet{ortega02} and \citet{song03}, nor is kinematics a reliable selector of a moving group \citep{song02}.  Most other indicators point to a $>$ 100~Myr age.  \citet{nordstrom04} determined an age of 900~Myr, but this age is uncertain given their methodology--the lower limit for the age is 0, while the upper limit is
2.2~Gyr.  Ca~II H and K line indicators point to an age of 500~Myr \citep{silverstone00}.  Although not an accurate age indicator for early F-type stars, the Lithium 6708\AA\ feature shows an equivalent width of 40~m\AA\ (Song~2008, in prep.), which is consistent with $\sim$100~Myr Pleiades stars. As a comparison, previously recognized 30~Myr old (Tucana-HorA) stars or younger early F-type stars all show equivalent widths for Li of $\sim$100m\AA.  The evidence to date points to an older age for HD~15115.

HD~15115 is a prime example of the potential complexity of nearby circumstellar debris disks.  It possesses a strange brightness asymmetry in its outer reaches that extends into about 2\arcsec (90~AU).  This asymmetry, however, is a function of wavelength, and at least part of the cause may be compositional or grain population differences between the two sides of the disk.  KFG07 investigated whether the close passage of a star might explain the strange morphology of the disk.  They found a candidate object, HIP~12545, that might have passed very close to the debris disk of HD~15115, but while its motion is consistent with being a co-moving object, backtracking of the the two star's positions back in time based on their Hipparcos/Tycho-2 astrometry and radial velocities shows that they are more widely separated in the past than they are now and could not have interacted.

The fact that we measure a warp only in the western part of the disk around 1\arcsec\ suggests that HD~15115 has an even more complex structure.  It would be
useful to confirm the presence of this structure in the disk at other wavelengths.  This is not easily done in the visible, unless STIS is refurbished in the next servicing mission--it is the only instrument able to image to within $\sim$1\arcsec\ in the visible.  Follow-up H and K (or F160W or F205W images with HST) will be useful to look for the warp as well as constrain the ultimate nature of the composition of the grains in the disk.  If water ice is indeed present, it will show strong absorption at around 2 and 3.5\micron.  Therefore, K and L band images of the scattered light disk are cruicial.

What are the other possibilities to explain both the warp, the asymmetry, and the varying nature of the dust grains?  One possibility is that the asymmetry is caused by a relatively recent collision that has injected many particles recently into the disk. Such an event is rare given the collisional times of the parent bodies and the timescale for such an event to be evenly spread in azimuth.  Similar asymmetries have been seen in the mid-IR with $\beta$-Pictoris \citep{weinberger03}, and it is striking that HD~15115 shares many of traits with this well known disk.

\acknowledgements
We thank M. Fitzgerald and P. Kalas for helpful discussions regarding their Keck observations and HD 15115 in general.  This research is based on observations with the NASA/ESA Hubble Space Telescope which is operated by the AURA, under NASA contract NAS 5-26555. These observations are associated with program GO \#10540.  Support for program \#10540 was provided by NASA through a grant from STScI.  AJW also acknowledges support from the NASA Astrobiology Institute. 

\bibliography{scibib}

\begin{thebibliography}{20}
\expandafter\ifx\csname natexlab\endcsname\relax\def\natexlab#1{#1}\fi

\bibitem[{{Ardila} {et~al.}(2007){Ardila}, {Golimowski}, {Krist}, {Clampin},
  {Ford}, \& {Illingworth}}]{ardila07}
{Ardila}, D.~R., {Golimowski}, D.~A., {Krist}, J.~E., {Clampin}, M., {Ford},
  H.~C., \& {Illingworth}, G.~D. 2007, \apj, 665, 512

\bibitem[{{Debes} {et~al.}(2008){Debes}, {Weinberger}, \&
  {Schneider}}]{debes08}
{Debes}, J.~H., {Weinberger}, A.~J., \& {Schneider}, G. 2008, \apjl, 673, L191

\bibitem[{{Decin} {et~al.}(2003){Decin}, {Dominik}, {Waters}, \&
  {Waelkens}}]{decin03}
{Decin}, G., {Dominik}, C., {Waters}, L.~B.~F.~M., \& {Waelkens}, C. 2003,
  \apj, 598, 636

\bibitem[{{Golimowski} {et~al.}(2006){Golimowski}, {Ardila}, {Krist},
  {Clampin}, {Ford}, {Illingworth}, {Bartko}, {Ben{\'{\i}}tez}, {Blakeslee},
  {Bouwens}, {Bradley}, {Broadhurst}, {Brown}, {Burrows}, {Cheng}, {Cross},
  {Demarco}, {Feldman}, {Franx}, {Goto}, {Gronwall}, {Hartig}, {Holden},
  {Homeier}, {Infante}, {Jee}, {Kimble}, {Lesser}, {Martel}, {Mei},
  {Menanteau}, {Meurer}, {Miley}, {Motta}, {Postman}, {Rosati}, {Sirianni},
  {Sparks}, {Tran}, {Tsvetanov}, {White}, {Zheng}, \& {Zirm}}]{golimowski06}
{Golimowski}, D.~A., {et~al.} 2006, \aj, 131, 3109

\bibitem[{{Hines} {et~al.}(2007){Hines}, {Schneider}, {Hollenbach}, {Mamajek},
  {Hillenbrand}, {Metchev}, {Meyer}, {Carpenter}, {Moro-Mart{\'{\i}}n},
  {Silverstone}, {Kim}, {Henning}, {Bouwman}, \& {Wolf}}]{hines07}
{Hines}, D.~C., {et~al.} 2007, \apjl, 671, L165

\bibitem[{{Kalas} {et~al.}(2007){Kalas}, {Fitzgerald}, \& {Graham}}]{kalas07}
{Kalas}, P., {Fitzgerald}, M.~P., \& {Graham}, J.~R. 2007, \apjl, 661, L85

\bibitem[{{Kalas} \& {Jewitt}(1995)}]{kalas95}
{Kalas}, P. \& {Jewitt}, D. 1995, \aj, 110, 794

\bibitem[{{Mo{\'o}r} {et~al.}(2006){Mo{\'o}r}, {{\'A}brah{\'a}m}, {Derekas},
  {Kiss}, {Kiss}, {Apai}, {Grady}, \& {Henning}}]{moor06}
{Mo{\'o}r}, A., {{\'A}brah{\'a}m}, P., {Derekas}, A., {Kiss}, C., {Kiss},
  L.~L., {Apai}, D., {Grady}, C., \& {Henning}, T. 2006, \apj, 644, 525

\bibitem[{{Nordstr{\"o}m} {et~al.}(2004){Nordstr{\"o}m}, {Mayor}, {Andersen},
  {Holmberg}, {Pont}, {J{\o}rgensen}, {Olsen}, {Udry}, \&
  {Mowlavi}}]{nordstrom04}
{Nordstr{\"o}m}, B., {et~al.} 2004,
  \aap, 418, 989

\bibitem[{{Ortega} {et~al.}(2002){Ortega}, {de la Reza}, {Jilinski}, \&
  {Bazzanella}}]{ortega02}
{Ortega}, V.~G., {de la Reza}, R., {Jilinski}, E., \& {Bazzanella}, B. 2002,
  \apjl, 575, L75

\bibitem[{{Schneider} {et~al.}(1999){Schneider}, {Smith}, {Becklin}, {Koerner},
  {Meier}, {Hines}, {Lowrance}, {Terrile}, {Thompson}, \&
  {Rieke}}]{schneider99}
{Schneider}, G., {et~al.} 1999, \apjl, 513, L127

\bibitem[{{Silverstone}(2000)}]{silverstone00}
{Silverstone}, M.~D. 2000, PhD thesis, AA(UNIVERSITY OF CALIFORNIA, LOS
  ANGELES)

\bibitem[{{Song} {et~al.}(2002){Song}, {Bessell}, \& {Zuckerman}}]{song02}
{Song}, I., {Bessell}, M.~S., \& {Zuckerman}, B. 2002, \aap, 385, 862

\bibitem[{{Song} {et~al.}(2003){Song}, {Zuckerman}, \& {Bessell}}]{song03}
{Song}, I., {Zuckerman}, B., \& {Bessell}, M.~S. 2003, \apj, 599, 342

\bibitem[{{Thompson} {et~al.}(1998){Thompson}, {Rieke}, {Schneider}, {Hines},
  \& {Corbin}}]{thompson98}
{Thompson}, R.~I., {Rieke}, M., {Schneider}, G., {Hines}, D.~C., \& {Corbin},
  M.~R. 1998, \apjl, 492, L95+

\bibitem[{{van Leeuwen}(2007)}]{vanl}
{van Leeuwen}, F. 2007, \aap, 474, 653

\bibitem[{{Weinberger} {et~al.}(2002){Weinberger}, {Becklin}, {Schneider},
  {Chiang}, {Lowrance}, {Silverstone}, {Zuckerman}, {Hines}, \&
  {Smith}}]{weinberger02}
{Weinberger}, A.~J., {et~al.} 2002, \apj, 566, 409

\bibitem[{{Weinberger} {et~al.}(2003){Weinberger}, {Becklin}, \&
  {Zuckerman}}]{weinberger03}
{Weinberger}, A.~J., {Becklin}, E.~E., \& {Zuckerman}, B. 2003, \apjl, 584, L33

\bibitem[{{Williams} \& {Andrews}(2006)}]{williams06}
{Williams}, J.~P. \& {Andrews}, S.~M. 2006, \apj, 653, 1480

\bibitem[{{Zuckerman} \& {Song}(2004)}]{zuckerman04}
{Zuckerman}, B. \& {Song}, I. 2004, \apj, 603, 738

\end{thebibliography}
\bibliographystyle{apj}

\end{document}